\documentclass[10pt,twocolumn,letterpaper]{article}

\usepackage[final]{cvpr}

\makeatletter
\DeclareRobustCommand\onedot{\futurelet\@let@token\@onedot}
\def\@onedot{\ifx\@let@token.\else.\null\fi\xspace}

\def\eg{\emph{e.g}\onedot}

\makeatother

\title{ADCNet: Learning from Raw Radar Data via Distillation}

\author{Bo Yang$^1$ \hspace{20pt} Ishan Khatri$^1$ \hspace{20pt} Michael Happold$^2$ \hspace{20pt} Chulong Chen$^1$\\
{\tt \small $^1$\{bo.yang, ishan.khatri, chulong.chen\}@motional.com \hspace{20pt} $^2$mhappold@startmail.com}
}

\begin{document}
\maketitle

\begin{abstract}
    As autonomous vehicles and advanced driving assistance systems have entered wider deployment, there is an increased interest in building robust perception systems using radars. Radar-based systems are lower cost and more robust to adverse weather conditions than their LiDAR-based counterparts; however the point clouds produced are typically noisy and sparse by comparison. In order to combat these challenges, recent research has focused on consuming the raw radar data, instead of the final radar point cloud. We build on this line of work and demonstrate that by bringing elements of the signal processing pipeline into our network and then pre-training on the signal processing task, we are able to achieve state of the art detection performance on the RADIal dataset. Our method uses expensive offline signal processing algorithms to pseudo-label data and trains a network to distill this information into a fast convolutional backbone, which can then be fine-tuned for perception tasks. Extensive experiment results corroborate the effectiveness of the proposed techniques.
\end{abstract}

\section{Introduction}
Perception systems for autonomous vehicles have seen significant advancements over the past decade. This progression is of considerable importance, given that these systems serve as the primary source of information for downstream tasks, including those responsible for motion prediction and planning. Consequently, the ability to perceive the environment accurately is of utmost importance for the effective functioning of an autonomous vehicle.


\begin{figure}[t]
    \centering
    \includegraphics[width=\linewidth]{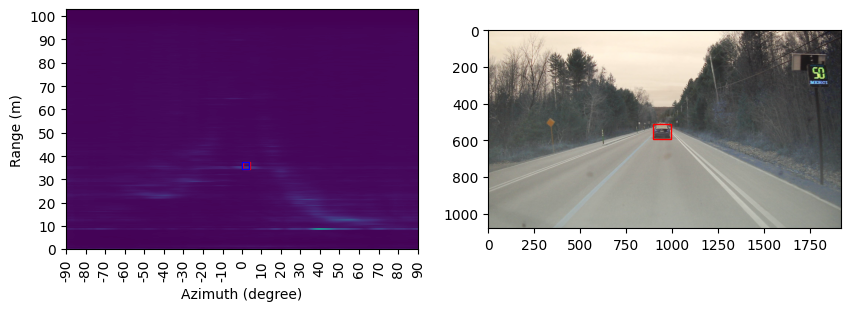}
    \includegraphics[width=\linewidth]{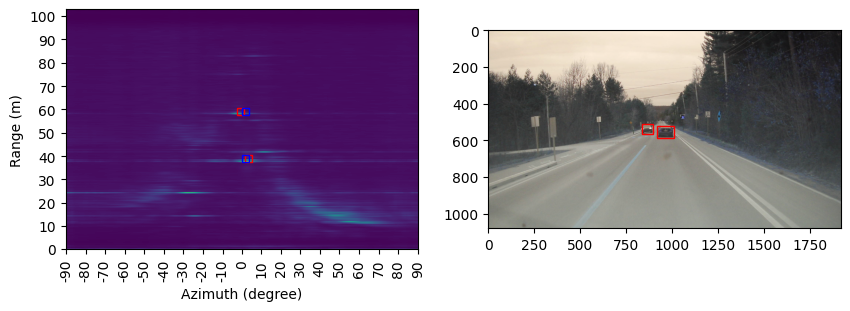}
    \includegraphics[width=\linewidth]{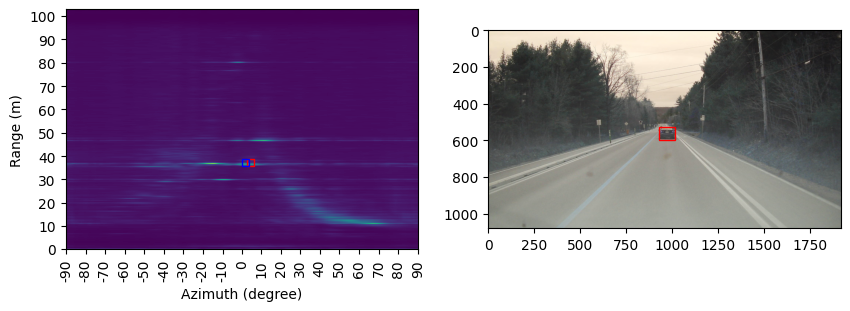}
    \includegraphics[width=\linewidth]{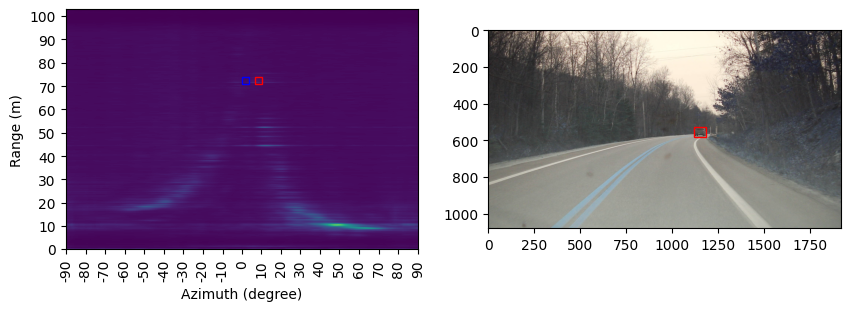}

    \caption{ADCNet prediction samples. The camera images are provided for reference only. The Range-Azimuth map is generated by the RADIal SDK. The small red squares in the RA image denote ground truth, while the blue squares denote network predictions }
    \label{fig:adcnet_samples}
\end{figure}

Radars have long been applied in automotive applications such as blind spot monitoring and adaptive cruise control, however, their applications to semantic scene understanding have been limited. This is due to the fact that traditional automotive radars' resolution is usually not enough to produce high quality sensing output, rendering them insufficient for scene understanding tasks such as object detection.

In the past few years, imaging radars for autonomous driving have emerged. Compared to traditional automotive radars, imaging radars usually operate at around 77GHz, with roughly 4GHz available bandwidth -- a much larger band than that of traditional automotive radars. In addition, these imaging radars are often equipped with multiple transmit and receiving antenna. Utilizing Multiple-Input-Multiple-Output (MIMO) technology \cite{rao2018mimo}, these new radars achieve much improved range and angle resolution. 

Capitalizing on this improved sensing capability of imaging radar, several works \cite{meyer2021graph, hwang2022cramnet,  wang2021rethinking, li2022modality, qian2021robust, li2022exploiting, palffy2020cnn} have attempted building perception models on these radars. One common theme of these works is that perception is done with low-level radar data, instead of the radar detection data as in traditional automotive radars.
Low-level radar data refers to any radar sensor data that have not gone through the peak detection \& thresholding step of the signal processing chain (see Figure~\ref{fig:sp_radar}). There are several low-level radar data representations: range-doppler (RD) radar cube, range-azimuth-doppler (RAD) cube,  etc., that have gone through various stages of signal processing, but not the final peak detection step that leads to very sparse radar detections.

{However, as described in \cite{rebut2022raw}, using the RAD tensor can be costly for imaging radars, as advanced angle finding operations are difficult to implement in real-time systems. As a result, FFTRadNet \cite{rebut2022raw} uses the RD cubes, which contains no explicit angular information - a necessary component for 3D perception. This limitation motivates our pre-training via distillation pipeline. To equip the neural network with angle finding capability, we pre-train the network on the task of mimicking the full signal processing chain (from ADC to RAD) via a distillation pipeline.
Our method employs offline signal processing algorithms to generate the RAD cubes for each ADC sample as pseudo-labels. Importantly, using these pseudo-labels enables training the model on a large set of data as no human annotation is needed.
Training a neural network with these pseudo-labels distills these complex algorithms into the weights of the neural network, which can be run efficiently on GPU hardware.
}

In order to facilitate using the ADC to RAD conversion task as a pre-training step, we introduce a learnable signal processing module to the network. This module aims to mimic the first two steps of classic signal-processing pipelines. Our design jump-starts the training: since the raw ADC data is inherently intricate, learning from a generic model architecture can be immensely difficult, as shown in Sec.~\ref{sec:init_exp}.

In summary, we propose a framework which features 1) a pre-training pipeline that infuses angle finding capability into neural networks and 2) a learnable signal processing module that is tailored to radar signal. The proposed method is performant as it achieves SOTA performance on the RADIal dataset, while also being flexible in accommodating different backbones (e.g. to achieve better latency) or different offline signal processing algorithms as teachers. 

\section{Background on radar signal processing}
\begin{figure}[ht]
    \centering
    \includegraphics[width=0.9\linewidth]{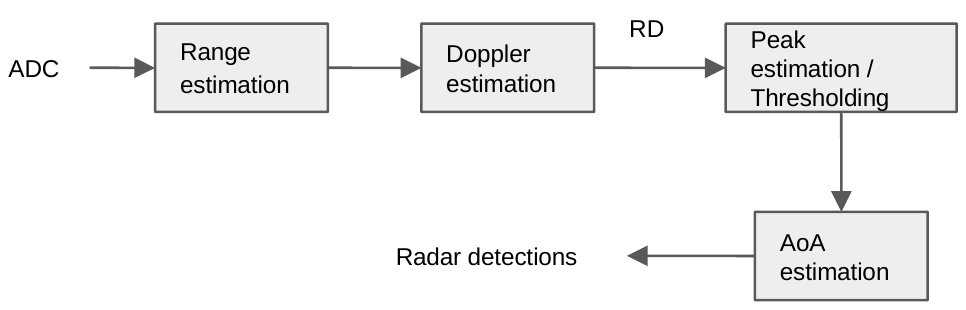}
    \caption{A simplified radar signal processing chain}
    \label{fig:sp_radar}
\end{figure}

This work is mainly concerned with a type of automotive radar termed Frequency Modulated Continuous Wave (FMCW) radar, see \eg \cite{brooker2005understanding}. At each measuring cycle, these radars send out a series of rapid ``chirps'' -- short waves with increasing frequency. At the receiving antenna, the reflections are captured, and sampled by an ADC device.
This digitized signal is then passed to other software modules in the radar sensor for signal processing.  A typical radar signal processing chain is shown in Figure~\ref{fig:sp_radar}.

\begin{figure}[ht]
    \centering
    \includegraphics[width=\linewidth]{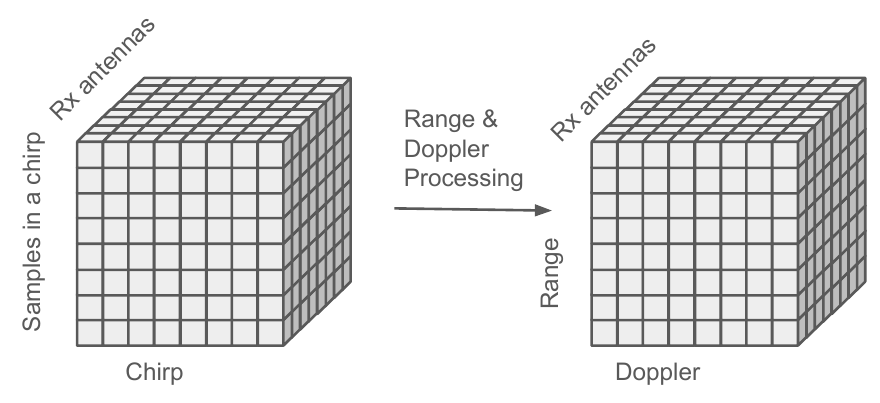}
    \caption{Illustration of ADC and RD arrays. Left: ADC, Right: RD. Notice the different dimensions between the two.}
    \label{fig:adc_rd_array}
\end{figure}
 The ADC data is usually organized as arrays as shown in Figure~\ref{fig:adc_rd_array}. Each frontal slab has dimension (number of samples per chirp, number of chirps), and holds all the samples for one receiving antenna. Signal processing operations (mainly DFTs) transform these two dimensions into range and Doppler dimensions, forming the RD array, as shown in the right part of Figure~\ref{fig:adc_rd_array}.

The next step of the SP chain is typically to threshold the RD cube in order to reduce the number of range-doppler bins that must be processed and then to run an angle-of-arrival (AOA) estimation algorithm. A rudimentary estimate of AOA can simply be found by adding a 3rd DFT along the final axis of the tensor, however in order to achieve higher quality estimates, iterative optimization algorithms such as IAA \cite{roberts2010iterative}, MUSIC \cite{van2002optimum} and ESPRIT \cite{roy1989esprit} can be used. While these algorithms have many advantages for automotive use (such as strong robustness \cite{sun2020mimocomparison}), they are difficult to run in real-time and require knowledge of the radar's parameters in order to implement \cite{li2023aess}.



\section{Related works}
\textbf{Radars for AV perception.} {There has been a considerable amount of work} utilizing radar detections for AV perception \cite{yang2020radarnet, shah2020liranet, nabati2021centerfusion}. In these papers, the radar detections are usually fused with another sensor modality, such as camera and LiDAR. The main reason is that radar detections are very sparse, and reliable semantic understanding of the driving scene is hard to achieve with radar alone.
In \cite{yang2020radarnet}, the authors propose a fusion scheme that combines radar detections and LiDAR point clouds, and obtain improved performance for object detection and velocity estimation. In \cite{shah2020liranet}, the authors combine radar detections and LiDAR point clouds and achieve improved performance for trajectory prediction, compared to using LiDAR alone.

There is a growing body of work attacking the AV perception problem by means of low-level radar data \cite{meyer2021graph, hwang2022cramnet,  wang2021rethinking, li2022modality, qian2021robust, li2022exploiting, palffy2020cnn}. CramNet \cite{hwang2022cramnet} proposes a method for fusing radar Range-Azimuth (RA) images with camera images using the attention mechanism. In \cite{meyer2021graph}, a graph convolution network is developed to work with radar RA data. In \cite{qian2021robust}, a radar-LiDAR fusion method is developed to improve perception robustness against adverse weather. In \cite{li2022modality}, the authors propose a method to cope with the missing modality problem for a radar-LiDAR fusion system. EchoFusion \cite{liu2023echoes} proposes a low-level radar and camera fusion scheme, where an attention mechanism using the polar coordinates is designed.

It is also worth highlighting the low-level radar datasets that have been facilitating the research in this direction. These include RADIATE \cite{sheeny2021radiate}, RADIal \cite{rebut2022raw}, ORR \cite{RadarRobotCarDatasetICRA2020}, CRUW \cite{Wang_2021_WACV}, while new ones (\eg K-Radar\cite{paek2022k}) continue to emerge. Different datasets often use different radar sensors, so their radar data representations tend to be different. For example, RADIATE provides radar radio-frequency (RF) images, ORR provides radar intensity maps, CRUW provies RA maps, while RADIal provides both ADC and Range-Doppler (RD) data. Since we are interested in exploring ADC data, we adopt the RADIal dataset to conduct our experiments, as among the currently ready-to-use radar datasets, RADIal is the only one that provides radar ADC data.

\textbf{Pre-training} Pre-training has been proven quite successful in several important domains, such as computer vision \cite{oord2018representation,he2020momentum,pmlr-v119-chen20j}, natural language understanding \cite{devlin2018bert, brown2020language}, speech and audio processing \cite{chung2020generative, li2021contrastive, zhang2022bigssl}, etc. These methods usually entail designing a surrogate learning task, such as predicting masked part of the input signal, so that the neural network learns intrinsic structure of the data. While deceptively simplistic, many works in this genre have shown that powerful representations, that support impressive performance on downstream tasks (such as image classification, question answering with natural language, automatic speech recognition), can be learned from a large collection of data.

\textbf{Distillation} Distillation and pseudo-labeling techniques are an important way to scale machine learning methods in regimes where human labels are too costly or too difficult to obtain. In MODEST \cite{you2022modest} a pipeline was created where LiDAR detection models were successively trained first with heuristic generated seed labels, and then on pseudo-labels generated by the previous iteration of the model. Distil-whisper \cite{gandhi2023distilwhisper} creates a high quality pseudo-labeled dataset for speech recognition. Finally, in ZeroFlow \cite{vedder2023zeroflow} the authors propose a similar offline optimization based pseudo-labeling technique for scene flow estimation, demonstrating the power of distilling offline techniques into small models by using larger unlabeled datasets.

\textbf{Learnable signal processing (SP) module} The learnable signal processing module idea for radar has been pursued in \cite{zhao2023cubelearn} for the hand gesture recognition task. Compared to \cite{zhao2023cubelearn}, our method uses \textit{perturbed} discrete Fourier transform (DFT), while \cite{zhao2023cubelearn} uses only exact DFT for initialization. Perturbation is an important novelty of this work, see discussions in Section~\ref{sec:learnable_sp} and experiment results in Table~\ref{tab:init}.

\begin{figure*}[ht]
\begin{center}
\includegraphics[width=\linewidth]{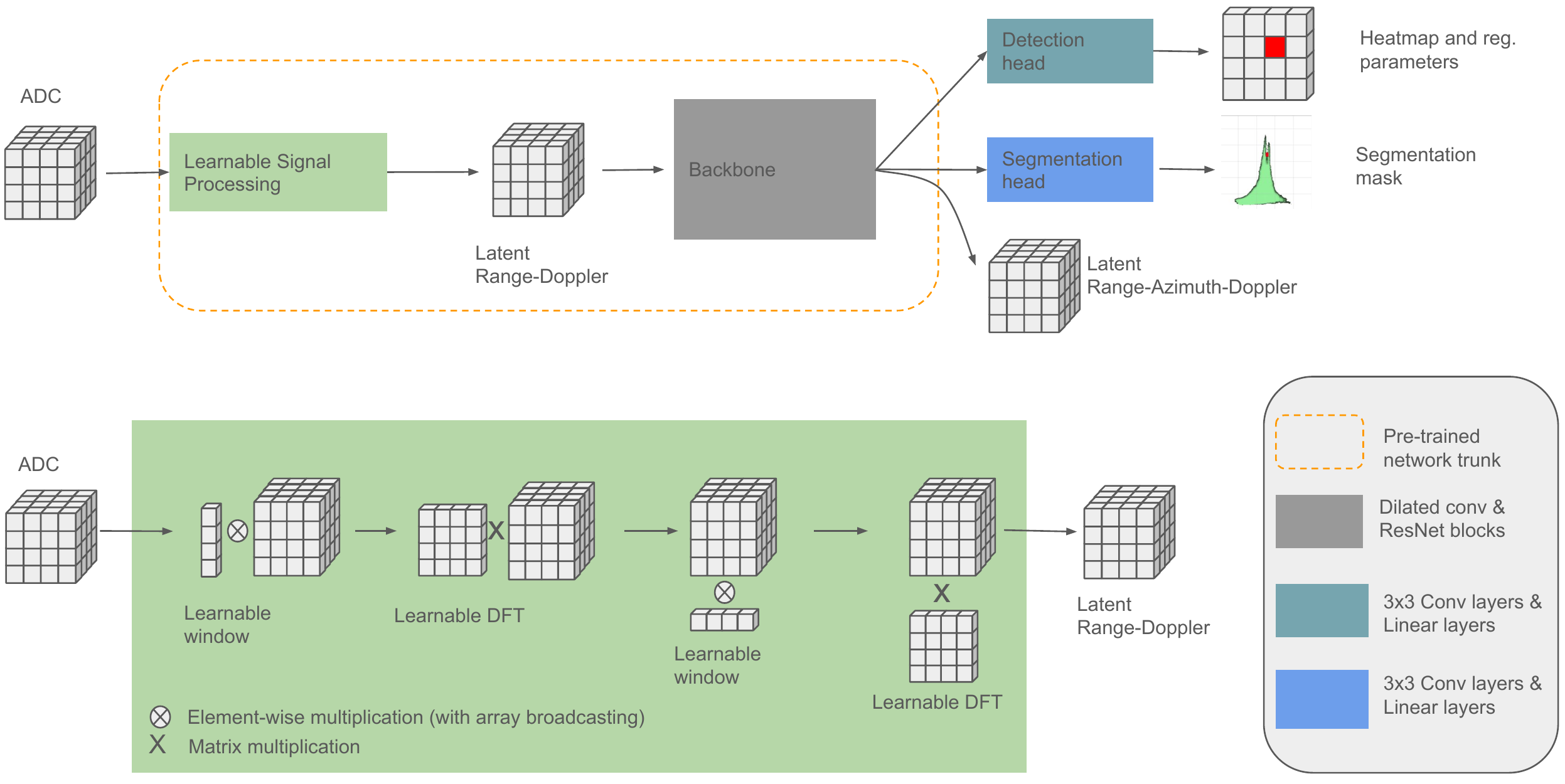}
\end{center}
   \caption{The ADCNet framework. A learnable signal processing module, as shown in the dashed box, is introduced at the start the network to enable end-to-end learning. The dashed yellow box represents the portion of the network that is pre-trained using our Distillation method described in Section \ref{sec:pre_training}}
\label{fig:adcnet}
\end{figure*}

\section{The RADIal dataset}\label{sec:radial}
To assist discussion, we provide a brief introduction to the RADIal \cite{rebut2022raw} dataset, and readers are referred to the original publication for more details.

The RADIal dataset provides synchronized radar, LiDAR and camera images for about 2 hours of driving. Labels for vehicle detection and freespace segmentation are also provided. In this work, we are interested in the radar data. For this, the dataset provides ADC data and RD data. The RADIal SDK \footnote{https://github.com/valeoai/RADIal} also enables generating RAD maps from the ADC data. 


The radar sensor used in the RADIal dataset is an imaging radar with 12 transmit antennas and 16 receiving antennas. The radar is mounted at the front of the vehicle, and has an approximate field-of-view (FOV) of 180 degrees, and a range of 103 meters.

The RADIal dataset contains a relatively balanced distribution of driving scenes in city, country side, and highway. For the object detection task and the freespace segmentation task, the labels are generated by models running on images and LiDAR, followed by human verification. Only vehicles are labeled for object detection.

Of the 25,000 time synchronized data frames, only about 8300 frames have object and segmentation labels. In Section~\ref{sec:pre_training} we will propose a method to utilize the unlabeled dataset for pre-training.

\section{The ADCNet Framework}
In the traditional radar signal processing chain, raw ADC data is converted into a Range Doppler cube via two DFT operations along two dimensions. The RD cube is then thresholded and run through an angle-of-arrival layer to estimate the azimuth of the targets. As explained in \cite{rebut2022raw}, this angle-finding step is typically the most computationally expensive process of the signal processing chain. The network used in \cite{rebut2022raw}, FFTRadNet uses RD cubes as input and then implicitly estimates the azimuth of the targets using the supervisory signal from the detection and segmenatation heads. In our work, we aim to bring the full signal-processing chain into the network and start from the raw ADC data. The goal is to design a framework for using raw ADC data and pre-training a network backbone that could be used for any number of downstream tasks (such as detection or segmentation).

Our framework is illustrated in Figure \ref{fig:adcnet}. Starting with the raw ADC data we run a learnable signal processing layer and then a backbone network. This network is trained to reconstruct the full RAD tensor that the classical signal processing pipeline would produce, using the SP pipeline as a teacher which provides ground-truth. In effect, we distil the full SP pipeline (which includes DFT, thresholding and AOA estimation) into the learnable SP module and the backbone. Once our backbone has been pre-trained we can fine tune it for downstream tasks such as detection and segmentation. Importantly, our framework is generalizable and can be applied to any sensor, signal processing chain, and backbone network architecture.

\subsection{Pre-training ADC to RAD via Distillation}\label{sec:pre_training}
The pre-training via distilation step of our framework is our most important contribution in this work. This pre-training can be roughly broken down into two steps.
\begin{enumerate}
    \item Using an offline SP algorithm, generate RAD from all the ADC data;
    \item Using the collected (ADC, RAD) pairs as input and (pseudo) labels to train the learnable SP and backbone network in a supervised learning fashion.
\end{enumerate}

Using the ADC as input, our learnable SP module + backbone network predicts RAD tensors which are supervised using the loss function described below:
\begin{align}
    \mathcal{L} = \text{smooth-L1}(\mathbf{Y}_{\text{RAD}} -  \hat{\mathbf{Y}}_{\text{RAD}}),
\end{align}
where $\mathbf{Y}_{\text{RAD}}$ denotes the generated RAD tensor and $\hat{\mathbf{Y}}_{\text{RAD}}$ denotes the prediction of the neural network. From the RADIal SDK, the generated RAD tensor is of shape $512 \times 751 \times 256$. We downsample the tensor to have shape $128 \times 248 \times 256$ for easier training. 
After pre-training the trunk of the ADCNet, we proceed to fine-tune the network with the multi-task setting as described in Section~\ref{sec:supervised}.

While conceptually simple, this step has a number of key advantages. The first of which is that it explicitly trains the network on the task of azimuth estimation which directly leads to an improvement in detection and segmentation performance (see Section \ref{sec:pretraining_exp}).

The second important advantage is that this step is a low cost addition to any existing pipeline. The data required does not need any human annotation which is a key advantage for raw radar datasets. This is because raw radar data is not human interpretable and thus requires sensor fusion techniques for accurate 3D labels, as shown in \cite{paek2022k}.

In this work we use the signal processing pipeline that is provided by the RADIal SDK to work with the specific radar sensor used to collect the RADIal dataset. The SDK provides a matched-filer type of algorithm for angle estimation. Clearly, more advanced algorithms (such as IAA, MUSIC etc.) could be used to generate the RAD with greater accuracy.  
However we are unable to try these alternative teacher algorithms for comparison due to lack of key radar configuration parameters (e.g. antenna array placement, multiplexing schemes, etc.). Investigation for better teacher algorithms is thus left as a future work.

\subsection{Learnable Signal Processing Layers}\label{sec:learnable_sp}
Since the DFT operations and windowing operations are relatively cheap, we can easily incorporate them into our network. This is similar to the method used in \cite{zhao2023cubelearn}, where the DFT is incorporated as a learnable module in the network, however unlike \cite{zhao2023cubelearn}, we also add the windowing function and perform a perturbed initialization described below. Both the DFT and windowing operations can be represented by matrix multiplications and we can simply set up linear layers which are initialized to the correct parameters in the network (see the Appendix for additional implementation details). However, we find that in practice initializing the SP module to use the exact DFT parameters can hurt performance, since the training process may fail to drive the weights away from the local minima to which they are initialized.  To alleviate the issue, we developed a novel perturbed-DFT initialization strategy: 
\begin{align}
    \mathbf{M} = \mathbf{M}_{\text{DFT}} + \mathcal{N}(0, \gamma)
    \label{eq:perturb}
\end{align}
In Eq.~\ref{eq:perturb}, $\mathbf{M}_{\text{DFT}}$ denotes the DFT matrix, and $\mathcal{N}(0, \gamma)$ denotes a random Gaussion matrix (with the same shape as the DFT matrix), with i.i.d. elements each has mean 0, and variance $\gamma$.  Intuitively, this $\gamma$ parameter should be small, such that the final matrix $\mathbf{M}$ used for initializing the SP module is perturbed, but still resemble the DFT matrix. In our experiments, the $\gamma$ is set to 0.1. The effect of this perturbation is examined in Section~\ref{sec:init_exp}

\subsection{Fine Tuning for Downstream Tasks}\label{sec:supervised}
To facilitate comparison with FFT-RadNet, we adopt the same multi-task learning setup. Specifically, our ADCNet backbone is fine-tuned to perform both object detection and freespace segmentation at the same time by adding detection and segmentation heads as shown in Figure \ref{fig:adcnet}.

For model training, the object detection part of the loss is
\begin{align}
    \mathcal{L}_{\text{det}} = \text{focal}(\mathbf{y}_{\text{cls}}, \hat{\mathbf{y}}_{\text{cls}}) + \alpha \text{smooth-L1}(\mathbf{y}_{\text{reg}}, \hat{\mathbf{y}}_{\text{reg}}),
\end{align}
where $\mathbf{y}_{\text{cls}}$ denotes the ground-truth classification labels on the feature map, with $1$ indicates presence of object while $0$ otherwise. The shape of  $\mathbf{y}_{\text{cls}}$ is $N_{\text{range-bins}} \times N_{\text{azimuth-bins}}$.The $\mathbf{y}_{\text{reg}}$ denotes the regression targets. Following \cite{rebut2022raw}, the network predicts a range and an azimuth value. The regression targets are thus the remainders of range and azimuth modulo range and azimuth bin sizes. The shape of $\mathbf{y}_{\text{reg}}$ is $N_{\text{range-bins}} \times N_{\text{azimuth-bins}} \times 2$, where the last dimension corresponds to range and azimuth. The $\text{focal}()$ function is from \cite{lin2017focal}, and the $\text{smooth-L1}()$ function is also known as Huber loss \cite{huber1992robust}. The $\alpha$ is a hyper-parameter balancing the two parts.

The freespace segmentation part of the loss is
\begin{align}
    \mathcal{L}_{\text{seg}} = \sum_{r, a} \text{BCE}(\mathbf{y}_{\text{seg}}(r, a), \hat{\mathbf{y}}_{\text{seg}}(r, a)),
\end{align}
where $\mathbf{y}_{\text{seg}}(r, a)$ denotes the ground truth at location $(r, a)$, with $1$ denotes object presence and $0$ otherwise. The $\text{BCE}()$ term denotes the binary cross entropy loss.

Combining the two parts, the loss function for training is thus
\begin{align}
    \mathcal{L} = \mathcal{L}_{\text{det}} + \beta\mathcal{L}_{\text{seg}},
\end{align}
where $\beta$ is a hyper-parameter controlling the weights on the freespace segmentation task.

\begin{table*}[!htp]\centering
\caption{Comparing ADCNet with baselines. AP: Average Precision, AR: Average Recall, RE: Range Error on detected objects, AE: Azimuth Error on detected objects. The full testset is broken down into an Easy and a Hard subset for detailed comparison. *: results cited from \cite{rebut2022raw}.}\label{tab:main_res}
\scriptsize
\resizebox{\textwidth}{!}{
\begin{tabular}{lrrrrrr|rrrrr|rrrrrr}\toprule
\textbf{} &\textbf{} &\multicolumn{5}{c|}{\textbf{All}} &\multicolumn{5}{c|}{\textbf{Easy}} &\multicolumn{5}{c}{\textbf{Hard}} \\\cmidrule{3-17}
\textbf{} &\textbf{Radar Input} &\textbf{F1} &\textbf{AP} &\textbf{AR} &\textbf{RE (m)} &\textbf{AE (°)} &\textbf{F1} &\textbf{AP} &\textbf{AR} &\textbf{RE (m)} &\textbf{AE (°)} &\textbf{F1} &\textbf{AP} &\textbf{AR} &\textbf{RE (m)} &\textbf{AE (°)} \\\midrule
FFT-RadNet * &RD &0.89 &0.97 &0.82 &0.11 &0.17 &0.95 &0.98 &0.92 &0.10 &0.13 &0.76 &0.93 &0.65 &0.13 &0.26 \\
Pixor - PC * &Point cloud &0.48 &0.96 &0.32 &0.17 &0.25 &0.45 &0.99 &0.29 &0.15 &0.19 &0.55 &0.93 &0.39 &0.19 &0.33 \\
Pixor - RA * &RA &0.89 &0.97 &0.82 &0.10 &0.20 &0.92 &0.97 &0.88 &0.09 &0.16 &\textbf{0.81} &0.96 &0.70 &0.12 &0.27 \\
Cross Modal DNN \cite{Yi2023CrossModal}    &RD  &0.90 &0.97 &0.84 &-    &-    &0.95 &0.98 &0.92 &-    &-    &0.77 &0.93 &0.66 &-    &-    \\
T-FFTRadNet \cite{giroux2023tfftradnet}    &RD  &0.90 &0.90 &0.90 &0.15 &0.12 &- &- &- &- &- &- &- &- &- &- \\
T-FFTRadNet \cite{giroux2023tfftradnet}    &ADC &0.87 &0.88 &0.87 &0.16 &0.13 &- &- &- &- &- &- &- &- &- &- \\
ADCNet (Ours) &ADC &\textbf{0.92} &0.95 &0.89 &0.13 &0.11 &\textbf{0.97} &0.96 &0.98 &0.12 &0.11 &\textbf{0.81} &0.91 &0.73 &0.16 &0.12 \\
\bottomrule
\end{tabular}
}
\end{table*}

\begin{table}[!htp]\centering
\caption{Freespace segmentation comparison. The performance is measured by mIOU, the higher the value the better. *: results cited from \cite{rebut2022raw}.}\label{tab:freespace}
\scriptsize
\begin{tabular}{lrrrr}\toprule
&All &Easy &Hard \\\midrule
FFT-RadNet * &74.00\% &74.60\% &72.30\% \\
PolarNet \cite{nowruzi2021polarnet} * & 60.6\%& 61.9\% & 57.4\% \\
Cross Modal DNN \cite{Yi2023CrossModal}   &80.40\% &81.60\% &76.70\% \\
T-FFTRadNet - RD \cite{giroux2023tfftradnet}   &80.20\% &-\% &-\% \\
T-FFTRadNet - ADC \cite{giroux2023tfftradnet}   &79.60\% &-\% &-\% \\
ADCNet &78.59\% &79.63\% &75.90\% \\
\bottomrule
\end{tabular}
\end{table}

\section{Experiment results}
In this section, we detail experiment results to verify the effectiveness of the proposed techniques. We present an overall comparison in Section~\ref{sec:exp_overall}, with ablation studies in the following subsections.

\subsection{Experiment setup}
We use the same data partitioning as that of \cite{rebut2022raw}. The whole labelled dataset is split into train, val, and test set by \textit{sequence} (driving session). That is, data samples from a sequence can only appear in one of the splits. This is to avoid allocating closely-resembling data samples (\eg two consecutive data samples in the same driving session) to the same split, which can lead to inflated performance metrics.

For evaluation, we also follow the setup as in \cite{rebut2022raw}. Object detection performance is measured by average-precision, average-recall and average-f1 scores, where average is done across a range of detection thresholds. The freespace segmentation results are evaluated with mIOU metric: the average IOU (intersection over union) score across test samples.

For the pre-training experiments in Section~\ref{sec:pretraining_exp}, it is important to note that we make sure any samples from the labeled test set sequences are not included in the pre-training stage. This avoids giving unfair advantages to the ADCNet, so that it is not allowed to see test samples even in the pre-training stage.

\subsection{Comparing ADCNet to baselines}\label{sec:exp_overall}
The object detection performance results are shown in Table~\ref{tab:main_res}, and the freespace segmentation results are presented in Table~\ref{tab:freespace}. Overall, we can see that ADCNet is the best method on object detection, capturing state of the art results even when compared to other methods which use more complex architectures such as T-FFTRadNet \cite{giroux2023tfftradnet} which uses a Swin Transformer based backbone. Looking at the breakdown of results we can see that ADCNet provides a 5\% F1 improvement over FFTRadNet on hard samples and is able to increase recall by 8\%. Additionally, when comparing just the hard samples, ADCNet is tied for the top score with Cross Modal DNN \cite{Yi2023CrossModal} which also uses additional supervision (from camera information) during training.

On the semantic segmentation task we achieve competitive results, falling only 2\% short of Cross Modal DNN \cite{Yi2023CrossModal} which captures the best performance. This suggests that the visual cues from camera signal may be providing important value for segmentation. Fortunately, this camera supervision technique can be combined with our proposed, radar-only framework. Exploring these fusion techniques is left for future work.

We present several model prediction samples in Figure~\ref{fig:adcnet_samples}. It can be seen from the samples that the network is able to correctly recognize the cars while rejecting other bright spots in the RA images. In the second sample, the network misses one car, possibly due to the road sign on the right, which causes a bright spot in the RA map, at a similar range of the missed car.

\subsection{Network architecture ablation studies}
As described in section \ref{sec:pre_training} our framework for pre-training via distillation can be used with other network architectures. In order to validate this and to demonstrate the effectiveness of our learnable SP modules we perform two experiments; first we replace the FFTRadNet backbone with a simple UNet to create ADC UNet, and second we remove the learnable SP module and replace it with 3D convolutions to create Conv3D + FFTRadNet. Both variations are trained with and without pre-training in order to understand the effects of our pre-training step on various networks. Please see the appendix for more details on the architecture of the ADC Unet and Conv3D + FFTRadNet model.
Additionally we train an ADCNet without pre-training for comparison.

From Table~\ref{tab:backbone}, when removing pre-training, we see a reduction of detection F1 score of 1\%, 5\% and 10\% for the ADCNet, the ADC UNet, and the Conv3d+FFTRadNet, respectively. There is an additional 2\% to 4\% drop in mIOU for the freespace segmentation performance for these methods. We suspect that the improved effect of pre-training on the UNet backbone is because it lacks any of the specific radar modules such as MIMO Pre-encoder that are found in FFTRadNet \cite{giroux2023tfftradnet}. These modules are helpful due to the Doppler division multiplexing used in the specific radar for this dataset. Nonetheless, owing to its simplicity, the UNet backbone runs in half the time as the FFTRadNet backbone and also consumes half the VRAM (see Table \ref{tab:latency}). This demonstrates the power of our pre-training framework: it allows simpler network architectures to punch far above their weight class by taking advantage of unlabeled data. Lastly, we find that the Conv3d+FFTRadNet architecture performs the worst overall, demonstrating the importance of using the learnable SP module that introduces radar inductive bias.


\begin{table}[!htp]\centering
\caption{Comparing different backbones using our framework. The numbers represent the performance on the full test set. Models denoted with NPT (no pre-training) are ones that do not use pre-training via distillation.}\label{tab:backbone}
\scriptsize
\resizebox{\linewidth}{!}{
\begin{tabular}{lrrrrrrr}\toprule
&\textbf{F1} &\textbf{AP} &\textbf{AR} &\textbf{RE(m)} &\textbf{AE (°)} &\textbf{mIOU} \\\midrule
ADCNet          &0.92 &0.95 &0.89 &0.13 &0.11 &78.59\% \\
ADCNet - NPT    &0.91 &0.96 &0.87 &0.12 &0.10 &74.45\% \\
ADC UNet        &0.85 &0.88 &0.82 &0.18 &0.11 &77.16\% \\
ADC UNet - NPT  &0.80 &0.83 &0.77 &0.19 &0.10 &73.04\% \\
Conv3d + FFTRadNet       &0.47 &0.58 &0.39 &0.19 &0.33 &74.69\% \\
Conv3d + FFTRadNet - NPT &0.37 &0.58 &0.27 &0.16 &0.46 &72.53\% \\
\bottomrule
\end{tabular}
}
\end{table}

\begin{table}[!htp]\centering
\caption{Efficiency comparison. Batch size is 20 for throughput and memory measurements. *All measurements taken on an RTX 3090 except for \cite{Yi2023CrossModal} which was measured with unknown hardware.}\label{tab:latency}
\scriptsize
\resizebox{\columnwidth}{!}{
\begin{tabular}{lrrrr}\toprule
&Latency (ms) &Throughput (samples/sec) &Memory (GB) \\\midrule
FFTRadNet \cite{rebut2022raw} & 15.93 &68  &12.33 \\
Cross Modal DNN \cite{Yi2023CrossModal} &68.00* &- &- \\
T-FFTRadNet \cite{giroux2023tfftradnet} &20.00 &- &- \\
ADCNet    & 18.13 &58  &14.89 \\
ADC U-Net & 8.18  & 137 &7.45 \\
\bottomrule
\end{tabular}
}
\end{table}

\subsection{Effectiveness of distillation}\label{sec:pretraining_exp}
In this section, we are interested in examining the prediction accuracy of RAD tensor in the pre-training stage. {A good RAD prediction accuracy would signal a successful distillation.} This is a valid concern, as both the ADC tensor and the RAD are large: ADC is of shape $512 \times 256 \times 8$ while RAD is $128\times 248 \times 256$ for the RADIal dataset. It is unclear whether learning a mapping between these two high-dimensional vectors is feasible.

\begin{figure}[bt]
    \centering
    \includegraphics[width=\linewidth]{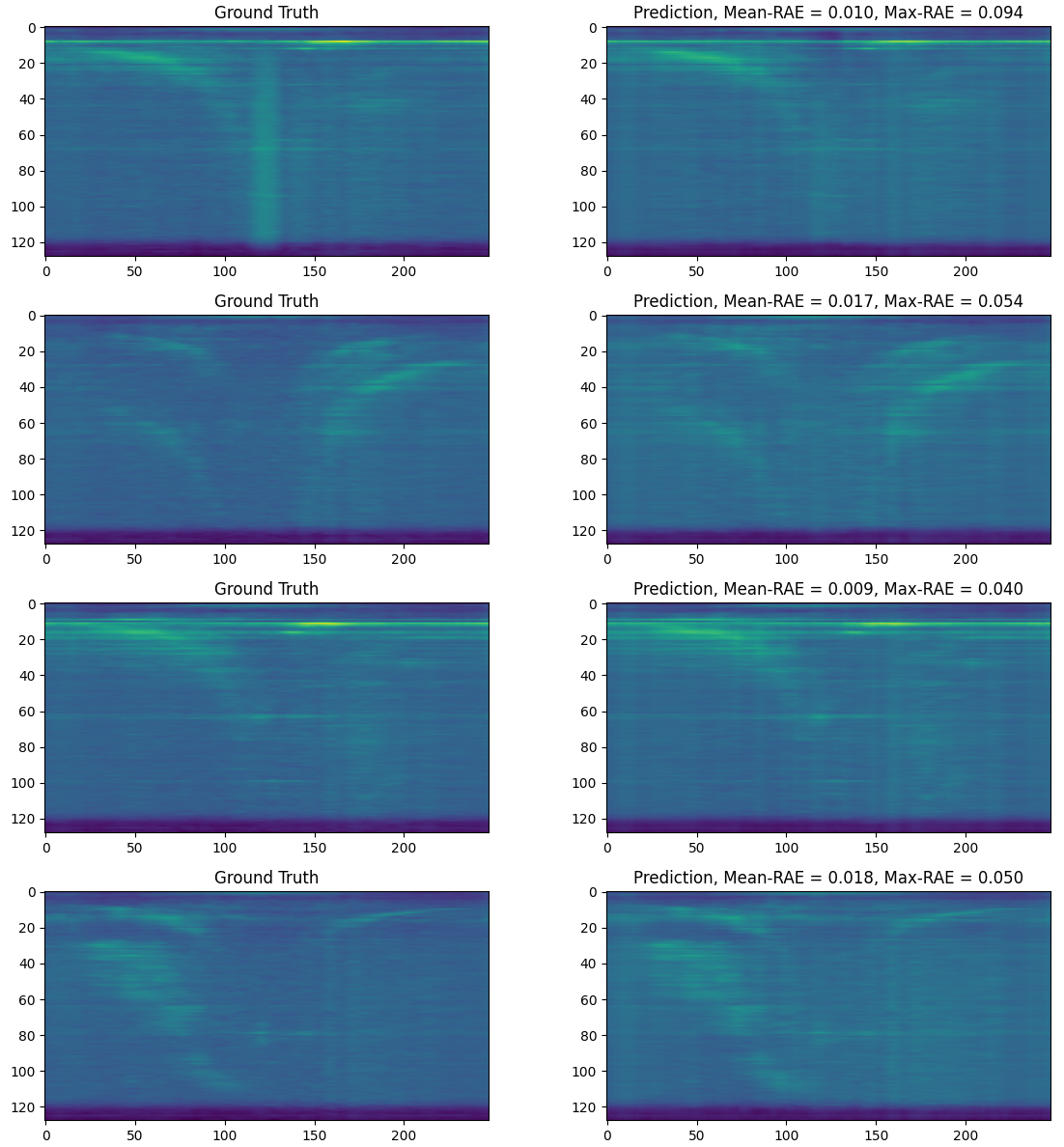}
    \caption{RAD prediction examples. The mean and max of Relative Absolute Error (RAE) for each sample are reported. As can be seen these examples, after pre-training, the network is able to predict RAD tensors to a very high accuracy.}
    \label{fig:rad}
\end{figure}

We visualize several randomly selected RAD prediction samples in Figure~\ref{fig:rad}. To visualize this 3D RAD tensor, we summed over the third dimension, and get a 2D image of dimension range and azimuth.
To get a quantitative error measure, we compute the Relative Absolute Error (RAE) for each RAD entry as 
\begin{align}
    \text{RAE}(i, j, k) = \frac{|\mathbf{Y}_{\text{RAD}}(i, j, k)  - \hat{\mathbf{Y}}_{\text{RAD}}(i, j, k) |}{|\mathbf{Y}_{\text{RAD}}(i, j, k) |}.
\end{align}
As such, we get a relative error measure for each entry, and we report the maximum and mean of these errors in Figure~\ref{fig:rad}.



\begin{table}[!htp]\centering
\caption{Comparing different initialization methods for ADCNet. The numbers represent the performance on the full test set.}\label{tab:init}
\scriptsize
\begin{tabular}{lrrrrrrr}\toprule
&\textbf{F1} &\textbf{AP} &\textbf{AR} &\textbf{RE(m)} &\textbf{AE (°)} &\textbf{mIOU} \\\midrule
Exact-DFT &0.87 &0.91 &0.84 &\textbf{0.12} &0.10 &\textbf{75.60\%} \\
Random-Doppler &0.82 &0.91 &0.74 &0.16 &0.10 &74.43\% \\
Perturbed-DFT &\textbf{0.91} &0.96 &0.87 &\textbf{0.12} &0.10 &74.50\% \\
Random &\multicolumn{6}{c}{Failed to converge} \\
\bottomrule
\end{tabular}
\end{table}

\begin{table*}[ht]\centering
\caption{Comparing Exact-DFT and Perturbed-DFT initialization method for the object detection task. Again NPT refers to the models that have not used pre-training via distillation.}\label{tab:perturbation_detection}
\scriptsize
\resizebox{\textwidth}{!}{
\begin{tabular}{lrrrrrr|rrrrr|rrrrrr}\toprule
\textbf{} &\textbf{} &\multicolumn{5}{c|}{\textbf{All}} &\multicolumn{5}{c|}{\textbf{Easy}} &\multicolumn{5}{c}{\textbf{Hard}} \\\cmidrule{3-17}
\textbf{} &\textbf{Initialization} &\textbf{F1} &\textbf{AP} &\textbf{AR} &\textbf{RE (m)} &\textbf{AE (°)} &\textbf{F1} &\textbf{AP} &\textbf{AR} &\textbf{RE (m)} &\textbf{AE (°)} &\textbf{F1} &\textbf{AP} &\textbf{AR} &\textbf{RE (m)} &\textbf{AE (°)} \\\midrule
ADCNet &Exact-DFT &0.89 &0.93 &0.86 &0.13 &\textbf{0.10} &0.96 &0.95 &0.97 &0.12 &0.10 &0.75 &0.86 &0.67 &\textbf{0.15} &0.12 \\
ADCNet &Perturbed-DFT &\textbf{0.92} &0.95 &0.89 &0.13 &0.11 &\textbf{0.97} &0.96 &0.98 &0.12 &0.11 &\textbf{0.81} &0.91 &0.73 &0.16 &0.12 \\
ADCNet - NPT &Exact-DFT &0.90 &0.96 &0.84 &0.13 &\textbf{0.10} &0.95 &0.98 &0.93 &0.12 &\textbf{0.09} &0.77 &0.91 &0.67 &\textbf{0.15} &0.12 \\
ADCNet - NPT &Perturbed-DFT &0.91 &0.96 &0.87 &\textbf{0.12} &\textbf{0.10} &\textbf{0.97} &0.98 &0.97 &\textbf{0.11} &0.10 &0.78 &0.91 &0.68 &0.16 &0.12 \\
\bottomrule
\end{tabular}
}
\end{table*}

\begin{table}[ht]\centering
\caption{Comparing Exact-DFT and Perturbed-DFT initialization method for the freespace segmentation task}\label{tab:perburbed_freespace}
\scriptsize
\begin{tabular}{lrrrrr}\toprule
& Initialization &\textbf{All} &\textbf{Easy} &\textbf{Hard} \\\midrule
ADCNet & Exact-DFT &77.28\% &78.37\% &74.50\% \\
ADCNet & Perturbed-DFT & 78.59\% &79.63\% &75.90\% \\
\midrule
ADCNet - NPT & Exact-DFT &70.39\% &70.81\% &69.31\% \\
ADCNet - NPT & Perburbed-DFT &74.45\% &75.85\% &70.89\% \\
\bottomrule
\end{tabular}
\end{table}

It can be seen from Figure~\ref{fig:rad} that the network can predict the RAD tensor to a very high accuracy: the ground truth and predicted images are visually close, and the RAE measures confirm this observation. This confirms that distillation of the expensive offline SP algorithm is indeed successful.

\subsection{Initialization for the learnable SP module}\label{sec:init_exp}
In this section, we compare different ways to initialize the learnable SP module. We compare these variations of the proposed ADCNet:
\begin{itemize}
    \item Perturb-DFT initialization: perturbed DFT matrices as shown in Equation~\ref{eq:perturb} are used to initialize both the range and Doppler DFT module in Figure~\ref{fig:adcnet}
    \item Exact-DFT initialization: the exact DFT matrices are used to initialize both range and Doppler DFT modules
    \item Random-Doppler initialization: the exact DFT matrix is used to initialize the range DFT, while a randomly generated matrix is used for Doppler 
    \item Random: randomly generated matrices are used for both range and Doppler modules
\end{itemize}
These methods employ a various amount of SP knowledge: the {Exact-DFT} method relies fully on SP, and the {Random} method almost completely disregard SP, while {Perturbed-DFT }and {Random-Doppler} lie in between.

The results are shown in Table~\ref{tab:init}. As can be seen, the {Perturbed-DFT } yields the best results. As we suspected, using the {Exact-DFT} initialization method can lead to sub-optimal performance. We hypothesize that this is because the network gets stuck in a local minima when initialized exactly. The {Random-Doppler} method yields considerably worse performance, while the {Random} method failed to converge to any meaningful model.

Since the Exact-DFT and Perburbed-DFT initialization methods get better performance than other choices, we provide a more in-depth comparison between the two. Moreover, we implement these initialization methods in the pre-training stage, then fine-tune to get the final detection and segmentation results. For this experiment, the multi-task learning setup as described in Section~\ref{sec:supervised} is performed, with and without pre-training. All the hyper-parameters are kept the same, and only the initialization method is varied.

The results are reported in Table~\ref{tab:perturbation_detection} and Table~\ref{tab:perburbed_freespace}. As can be seen from these tables, with and without pre-training, the Perturbed-DFT initialization method always yields better performance than Exact-DFT. The improvement by using Perturbed-DFT is especially visible for ADCNet, confirming our intuition that a larger dataset is needed for end-to-end learning, as limited training data may not be able to nudge the network too far from the SP-based initialization.

The $\gamma$ parameter in Equation~\ref{eq:perturb} should be chosen with care: a large $\gamma$ can destroy the DFT structure, while a too small $\gamma$ can leave the SP module in a non-optimal local minimum. Here we present an experiment, where different $\gamma$ is applied in the Perturb-DFT initialization scheme, and an ADCNet is trained (without pre-training) on the labeled RADIal dataset using the supervised multi-task learning setup in Section~\ref{sec:supervised}.
Table~\ref{tab:gamma} shows the results:  a large $\gamma$ (\eg $\gamma=2$) brings significant degradation to both object detection and freespace segmentation, while a small $\gamma$ like 0.1 achieves better performance than if no perturbation ($\gamma=0$) is applied.

\begin{table}[!htp]\centering
\caption{The effect of the $\gamma$ parameter in Perturb-DFT initialization}\label{tab:gamma}
\scriptsize
\begin{tabular}{rrrrrrrr}\toprule
$\gamma$ &\textbf{F1} &\textbf{AP} &\textbf{AR} &\textbf{RE(m)} &\textbf{AE (°)} &\textbf{mIOU} \\\midrule
0 &0.90 &0.96 &0.84 &0.13 &\textbf{0.1} &70.39\% \\
0.1 &\textbf{0.91} &0.96 &0.87 &\textbf{0.12} &\textbf{0.1} &\textbf{74.45\%} \\
0.5 &0.84 &0.88 &0.81 &0.16 &0.13 &70.03\% \\
2 &0.65 &0.75 &0.57 &0.18 &0.12 &69.47\% \\
\bottomrule
\end{tabular}
\end{table}


\section{Conclusion}
We present a pre-training via distillation pipeline that takes advantage of raw rdar data. By utilizing the task of ADC to RAD conversion as a pre-training step, we effectively distill important AOA estimation capabilities into our backbone network. Additionally, we propose a novel perturbed initialization technique for our improved learnable signal processing module. The combination of these techniques results in state of the art object detection and improved freespace segmentation results on the RADIal dataset within a unified multi-task model.

The designed pre-training via distillation process can be used with other SP algorithms as the teacher or other backbones as the student. We provide a simple example of replacing the backbone, and perform ablations to demonstrate the effectiveness of our pre-training pipeline as well as our perturbation initialization technique. We expect these techniques to be more powerful in more challenging environments (e.g. dense urban roads), or when scaling to larger unlabeled datasets for pre-training. We leave these explorations for future works.




\clearpage
\newpage
{
    \small
    \bibliographystyle{ieeenat_fullname}
    \bibliography{main}
}

\end{document}


\title{Supplementary for ``ADCNet: Learning from RAW Radar Data via Distillation''}

\maketitle

\section{More experiment details}
For training the ADCNet (both training from scratch and fine-turing from a pre-trained checkpoint), we use a single machine with 4  NVIDIA A10G GPUs. We use the Adam optimizer \cite{kingma2014adam}, with initial learning rate of $4\times 10^{-4}$, and a batch size of 4 per GPU. We use the sum of validation F1 score (for object detection) and validation mIOU score (for freespace segmentation) to pick the best checkpoint.

For pre-training, we use 4 nodes each equipped with 4  NVIDIA A10G GPUs. The learing rate $16 \times 10^{-4}$, and the batch size is 12. We train the model for a total of 60 epochs, and pick the model with smallest validation loss for the subsequent fine-tuning step.

\section{Learnable DSP training}

\begin{figure}[h]
    \centering
    \includegraphics[width=0.8\linewidth]{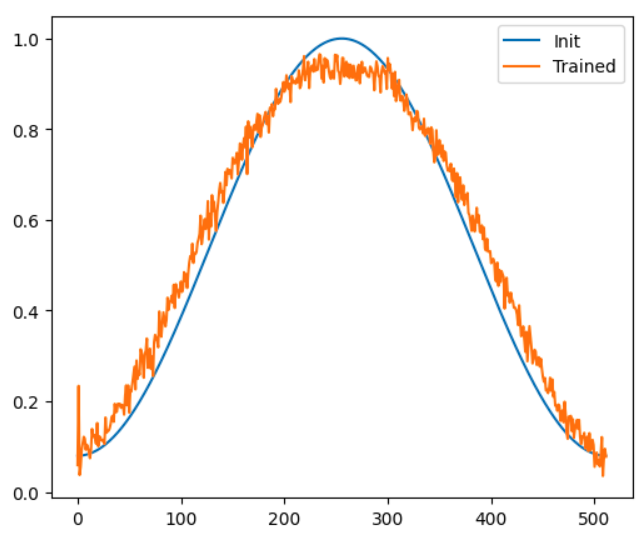}
    \caption{The first learnable window function of ADCNet before and after training}
    \label{fig:range_window}
\end{figure}

\begin{figure}[h]
    \centering
    \includegraphics[width=0.8\linewidth]{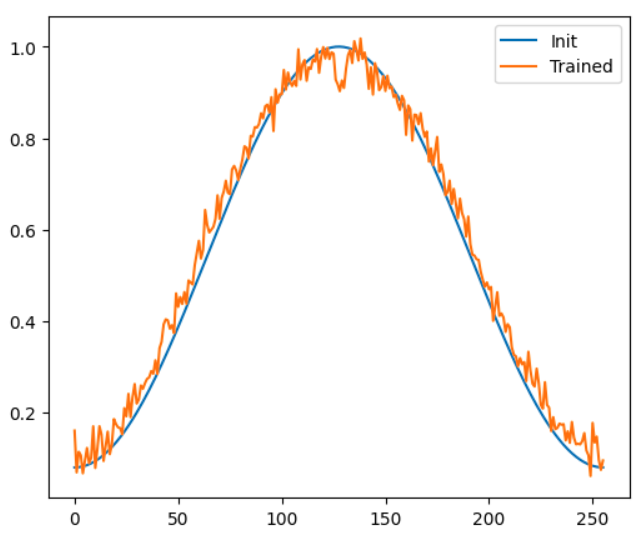}
    \caption{The second learnable window function of ADCNet before and after training}
    \label{fig:doppler_window}
\end{figure}

\subsection{Implementation of the learnable SP module}
\begin{figure}[th]
    \centering
    \includegraphics[width=\linewidth]{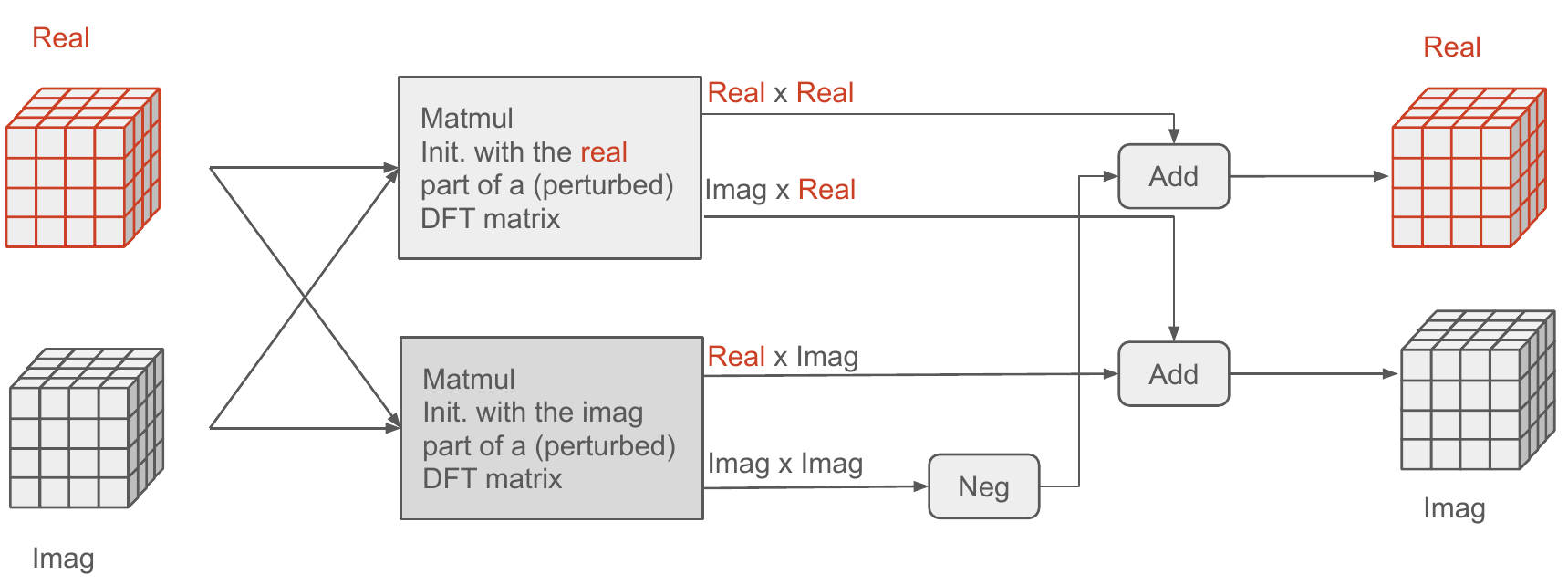}
    \caption{Implementing learnable DFT using real paramater matrices}
    \label{fig:learnable_dft}
\end{figure}

One difficulty of implementing the learnable SP module is the fact that the DFT operation is in complex domain. To avoid using complex operations in the neural network, we split the complex tensors (ADC array and DFT matrix) into real and imaginary parts, and perform the multiplications separately, as shown in Figure~\ref{fig:learnable_dft}. These are standard operations and can be easily implemented in a typical deep learning framework such as Pytorch. The learnable window module involves only one parameter vector, and it is multiplied to the inputs in the forward pass.

\subsection{Effects of training on the windowing functions}
The effect of end-to-end training on window functions are shown in Figure~\ref{fig:range_window} and Figure~\ref{fig:doppler_window}. It can be seen that in both the functions, the top is reduced and sides are raised. In signal processing, choice of window functions (see \eg \cite{window_funcs}) are usually made by a human expert, while the proposed approach amounts to data-driven window design -- a process that could be incorporated in radar signal processing implementation.

\begin{figure}[th]
    \centering
    \includegraphics[width=\linewidth]{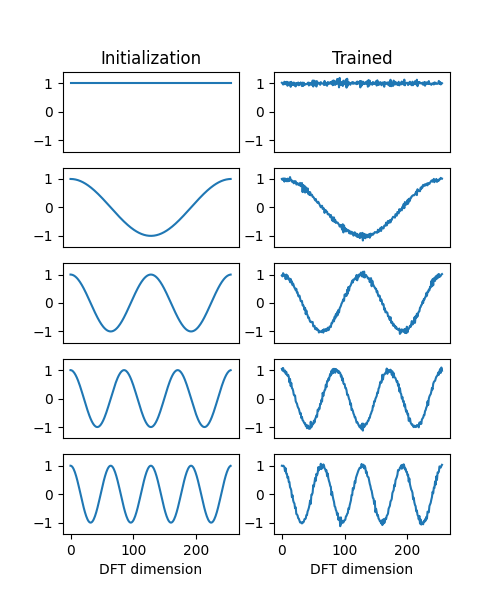}
    \caption{The real weights of the DFT matrix before and after training when \textit{Exact-DFT} initialization is used. x-axis: DFT index; y-axis: real part of the DFT coefficient.}
    \label{fig:exact_dft_training}
\end{figure}

\begin{figure}[th]
    \centering
    \includegraphics[width=\linewidth]{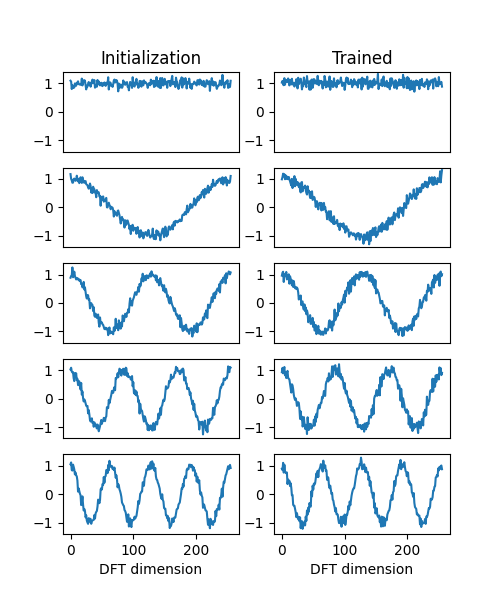}
    \caption{The real weights of the DFT matrix before and after training when \textit{Perturbed-DFT} ($\gamma = 0.1$) initialization is used. x-axis: DFT index; y-axis: real part of the DFT coefficient.}
    \label{fig:perturbed_dft_training}
\end{figure}

\subsection{Effects of perturbation on the learnable DFT}
When training with the \textit{Exact-DFT} initialization, we found that the DFT weights almost do not change before and after training, as shown in Figure~\ref{fig:exact_dft_training}. Figure~\ref{fig:exact_dft_training} shows the real part of the first 5 rows of the second DFT matrix in ADCNet. This confirms our suspicion that without perturbation, the learnable DFT layer can get stuck in the local optimum of the DFT matrix. For a comparison, the same visualization for ADCNet with \textit{Perturbed-DFT} initialization is shown in Figure~\ref{fig:perturbed_dft_training}.

\begin{table}[!htp]\centering
\caption{The difference between initialization and trained DFT weights, using either \textit{Exact-DFT} or \textit{Perturbed-DFT} initialization}\label{tab:init_diff}
\scriptsize
\begin{tabular}{lrrrrr}\toprule
&\multicolumn{2}{c}{The first learnable DFT} &\multicolumn{2}{c}{The second learnable DFT} \\\cmidrule{2-5}
&real &imaginary &real &imaginary \\\midrule
Exact-DFT &0.039 &0.039 &0.039 &0.039 \\
Perturbed-DFT &0.120 &0.120 &0.121 &0.121 \\
\bottomrule
\end{tabular}
\end{table}

To get a quantitative measure of the difference between the initialization and trained DFT weights, we compute the average absolute difference between the two set of weights. The real and imaginary part is computed separately, and the results are shown in Table~\ref{tab:init_diff}. As can be seen from Table~\ref{tab:init_diff}, with Exact-DFT initialization, the DFT matrix indeed does not change much compared with the one with Perturbed-DFT, confirming the necessity to add perturbation to the DFT matrix for a strong end-to-end model.

\section{Additional description on the ADC Unet and Conv3d+FFTRadNet baselines}
The ADC Unet and Conv3d+ FFTRadNet are designed such that they have similar number of learnable parameters as the ADCNet.

For ADC Unet, we use a standard Unet backbone to replace the backbone the in the ADCNet. Importantly, we also removed the dilated convolution layer that is supposed to exploit the radar signal structure. Removal of the dilated convolution layer is motivated by the fact that these special convolution layer runs much slower than the standard 3x3 convoutional layers. As a result, we get a baseline model that, albeit yields worse accuracy, runs much faster. The proposed pre-training technique improves this baseline quite notably (5\% absolute improvement on F1 score).

For the Conv3d+FFTRadNet baseline, we removed the learnable SP module from the ADCNet and replaced with a stack of Conv3d layers. The motivation is to test if such a generic model can handle the intricate radar signal purely from end2end training: the results from our experiment (in the main paper) is negative, highlighting the necessity of incorporating signal processing knowledge and techniques into the end2end learning regime. 
{
    \small
    \bibliographystyle{ieeenat_fullname}
    \bibliography{main}
}